# Surface-switchable nonreciprocity protected by Fermi arcs in Weyl semimetal TaAs

**Authors:** Dong Li[1]*, Yuki M. Itahashi[1], Ying-Ming Xie[1], Menghu Zhou[2], Yukako Fujishiro[1,3], Kunjie Zheng[4], Yao Guang[1], Max T. Birch[1], Ilya Belopolski[1], Max Hirschberger[1,5,6], Takahiro Morimoto[5], Xiao-Xiao Zhang[4], Zhi-An Ren[2], Naoto Nagaosa[1,7], Yoshihiro Iwasa[1]*

**Affiliations:**

[1]RIKEN Center for Emergent Matter Science; Wako, 351-0198, Japan

[2]Institute of Physics, Chinese Academy of Sciences; Beijing, 100190, China

[3]RIKEN Pioneering Research Institute; Wako, 351-0198, Japan

[4]Wuhan National High Magnetic Field Center and School of Physics, Huazhong University of Science and Technology, Wuhan 430074, China

[5]Department of Applied Physics, University of Tokyo; Tokyo, 113-8656, Japan

[6]Quantum-Phase Electronics Center, University of Tokyo; Tokyo, 113-8656, Japan

[7]RIKEN Fundamental Science Program, Wako, Saitama 351-0198, Japan.

*Corresponding author. Email: dong.li.hs@riken.jp (D.L.); iwasay@riken.jp (Y.I)

**Abstract:** Weyl semimetals host topologically protected surface states, known as Fermi arcs, which connect bulk Weyl nodes in momentum space[1,2]. Both bulk Weyl nodes and Fermi arcs are anticipated to be chiral[1-4]. The chirality of bulk bands has been confirmed through observations of the chiral anomaly[5,6] and Weyl orbits[7,8]. In contrast, despite their discovery more than a decade ago[9-13], the chiral nature of Fermi arcs has remained unresolved. Here we report Fermi-arc-induced nonlinear transport in the archetypal Weyl semimetal TaAs. Using focused ion beam techniques[14], we fabricated micro-scale devices that enable simultaneous transport measurements on opposing topological surfaces. While linear transport remains dominated by bulk conduction, nonlinear transport uncovers surface-specific contributions, including an exceptionally large third-order nonreciprocal response that exceeds conventional expectations and highlights the crucial role of the singular arc endpoints[15,16]. Our findings unambiguously demonstrate the chiral nature of Fermi arcs and establish nonlinear transport as a direct probe of these topological surface states. By revealing a surface-switchable, room-temperature nonlinear response that is topologically protected, this work introduces a new functionality in Weyl semimetals. Given the abundance of natural materials predicted to host topological semimetal states[17-19], these results open opportunities for exploring nonlinear transport phenomena and device concepts across a broad class of systems.

Topological materials are characterized by the emergence of robust surface states protected by the topology of the bulk band structure[20]. In topological insulators, these metallic surface states are readily distinguished from the insulating bulk and give rise to a broad family of Hall effects observed in transport measurements[21]. In contrast, detecting the topological surface states in Weyl semimetals, known as Fermi arcs[1], remains challenging due to the overwhelming contribution of metallic bulk states to charge transport. Fermi arcs arise from the nonzero momentum-resolved Chern number connecting bulk Weyl points and are theoretically predicted to reverse direction between the top and bottom surfaces[3,4]. The chirality embedded in the bulk electronic bands has been inferred from experimental observations of the chiral anomaly [5,6] and chirality transfer via the Weyl orbits[7,8]. Nevertheless, despite the clear evidence for the existence of Fermi arcs[9-11] and their associated spin textures[12,13], their expected chiral or unidirectional nature, which depends on the crystal symmetry, has yet to be resolved experimentally.

Definite chirality in quantum materials can be detected and distinguished through their nonlinear responses[22,23]. For instance, nonreciprocal chiral transport often manifests as nonlinear nonreciprocal magnetoresistance $R(I,B)=R_0(1+\beta B^2+\gamma IB)$ [22], where the resistance $R$ depends on the directions of applied current $I$ and external magnetic field $B$ due to a non-zero nonreciprocal coefficient $\gamma$, beyond the conventional even-in-field term $\beta B^2$. The nonreciprocal term $\gamma IB$ is known as electronic magnetochiral anisotropy[24] and emerges when spatial inversion ($\mathcal{P}$) is broken and time-reversal ($\mathcal{T}$) symmetry is effectively broken by an external magnetic field. Such a nonreciprocal response is inherently very weak[22], posing a challenge for nonlinear chiral transport measurements in bulk samples. Recent advances in focused-ion-beam (FIB) technology[14] have enabled the fabrication of micro-sized devices (Methods), thereby facilitating the observation of nonlinear transport in quantum materials, particularly those that cannot be mechanically exfoliated, such as TaAs. Moreover, the flexible three-dimensional design of FIB-fabricated devices enables simultaneous access to both the top and bottom topological surface states.

In this work, we demonstrate the chiral nature of Fermi arcs through surface-switchable nonlinear transport between the top and bottom surfaces of TaAs. Using FIB-fabricated devices, we observe quantum oscillations that confirm excitations of bulk Weyl fermions. In the linear response, the first-harmonic resistance $R^{1\omega}$ is nearly identical on opposite surfaces, as it is dominated by metallic bulk states. By contrast, in the nonlinear response, we detect sign-reversed second-harmonic resistance $R^{2\omega}$ between the top and bottom surfaces, which is our central experimental observation. These surface-switchable $R^{2\omega}$ signals persist up to room temperature and remain robust against FIB-induced surface damage, consistent with a topologically protected origin. Unexpectedly, $R^{2\omega}$ exhibits anomalous field and angular dependence, coinciding with the emergence of a third-order nonreciprocal response. These third-order components are highly reproducible across all ten devices measured. A theoretical model based on the conventional Boltzmann approach qualitatively captures the third-order term but fails to reproduce the experimental magnitudes quantitatively. The most plausible explanation lies in the singularity at the endpoints of Fermi arcs, where the quantum geometry diverges and becomes difficult to evaluate. Together, our results establish nonlinear transport as a powerful probe of Fermi arcs in Weyl semimetals and highlight the potential of surface-switchable nonlinear responses for future chiral electronic applications across a broad class of topological semimetals.

## Chirality and linear transport in TaAs

Tantalum Arsenide (TaAs, space group $I4_1md$) is a prototypical Weyl semimetal with broken $\mathcal{P}$ symmetry[25], hosting twelve pairs of Weyl nodes protected by mirror symmetries in the *ab*-plane. Figure 1a illustrates the two types of chirality in TaAs: chiral Weyl nodes in the bulk and and Fermi arcs with opposite orientations on the surfaces. The Fermi arcs appear exclusively on the top and bottom surfaces[2], exhibiting opposing Fermi velocities[3,4]. Figure 1b shows a schematic of our device design that allows simultaneous measurements on the top and bottom surfaces via the sidewalls. We fabricated TaAs devices with typical dimensions of $100\times50\times4$ $\mu m^3$ using FIB techniques. Figure 1c shows a top-view scanning electron microscope image of a representative TaAs device D1. Images of all devices are provided in Extended Data Fig. 1.

To verify the presence of Weyl fermions in our transport measurements, we conducted the Shubnikov–de Haas (SdH) measurements under $B$ applied along the *c*-axis (Fig. 1d). The SdH quantum oscillations reveals two Fermi pockets $F_\alpha = 6.2$ T and $F_\beta = 16.2$ T (Fig. 1e), with corresponding effective masses $m^* = 0.057\ m_e$ and $0.139\ m_e$, respectively (Extended Data Fig. 2). These pockets are consistent with the topological Weyl nodes ($F_\alpha$) and topologically trivial hole pockets ($F_\beta$) reported in previous quantum oscillation study on TaAs crystals[26]. The Weyl orbit is not observed in TaAs device D9 (thickness 450 nm; Fig. S1), the thinnest sample in this study. Its formation requires sample thicknesses shorter than the mean free path (~100 nm in TaAs[26]), presenting a significant challenges for device fabrication. Although the chiral anomaly has been reported in TaAs under parallel electric and magnetic fields ($E//B$)[5,6], we observe only positive magnetoresistance in both $I//B//c$-axis and $I//B//b$-axis configurations (Fig. 1f). Previous reports of low-field negative magnetoresistance were attributed to current-jetting artifacts[6,27], which are largely eliminated in our FIB-fabricated devices owing to the well-defined current confinement [28]. Furthermore, the chiral anomaly signal remains obscured under moderate magnetic fields in TaAs due to ultrahigh carrier mobility and non-Weyl background from hole-type quasiparticles[6]. These limitations of linear charge transport in probing bulk chirality TaAs, including both Weyl orbits and the chiral anomaly, underscore the need for alternative approaches to directly access the chiral topological surface states in Weyl semimetals.

## Surface-switchable nonlinear transport

We then performed nonlinear transport measurements on opposing topological surfaces to explore the properties of the Fermi arcs. To avoid artifacts in the nonlinear signals, we employed advanced FIB fabrication procedures while deliberately excluding FIB-deposited asymmetric platinum electrodes[29]. Nonlinear transport manifests itself as the $R^{2\omega}$ in AC transport measurements with a lock-in frequence $\omega$[30,31]. As described by $R^{2\omega}(J,B) \propto R_0\gamma' JB$, the signal scales linearly with magnetic field $B$, current density $J$, and a reduced nonreciprocal coefficient $\gamma' = \gamma A$, which is normalized by the sample cross section area $A$. By applying large current density ($J \sim 10^7$-$10^8$ A/m$^2$) in micro-sized devices, we enhanced $R^{2\omega}$ to detectable levels, enabling the long-sought-after observation of nonlinear transport in TaAs, which had eluded previous efforts. Figure 2a shows the magnetic field dependence of $R^{2\omega}$ measured on the top and bottom surfaces. Remarkably, the $R^{2\omega}$ signals exhibit opposite signs between the two surfaces. If $R^{2\omega}$ originated from bulk polarization, as TaAs possesses a polar axis along the *c*-axis, the signals on the top and bottom surfaces would have the same sign. However, surface polarization inherently reverses direction between opposing surfaces. Therefore, the observed sign reversal of $R^{2\omega}$ provides compelling evidence that the nonlinear signal arises from surface contributions rather than the bulk. This sign reversal in $R^{2\omega}$ between the top and bottom surfaces is reminiscent of the nonlinear transport observed in the chiral edge states of magnetic topological insulators[32]. Another notable feature is the non-monotonic magnetic field dependence of $R^{2\omega}(B)$, which saturates and even decreases at high fields. This anomaly persists across a range of current amplitudes (see Extended Data Fig.3) and is consistently observed in multiple devices, confirming its intrinsic origin. In contrast, the $R^{1\omega}$ signal measured on the top and bottom surfaces is nearly identical (fig. S2), indicating that linear transport predominantly probes bulk electronic states rather than surface contributions.

Figure 2b shows the current dependence of $R^{2\omega}(I)$ for both surfaces, exhibiting linearity in the low-current regime, with deviations appearing above 0.6 mA due to Joule heating effects. All nonlinear transport measurements were carefully conducted in the low-current linear regime to avoid interference from Joule heating. Figure 2c presents the angular dependence of $R^{2\omega}(\theta)$, where $\theta$ is the angle between the current between $I$ and $B$ within the *ab*-plane. At 0.5 T, the data are well described by a simple $\sin(\theta)$ dependence. However, at higher fields such as 2 T, the response deviates markedly from this form due to increasing nonlinearity with respect to $B$. This deviation indicates the emergence of additional contributions in the high-field regime that cannot be captured by a simple sinusoidal model. These anomalous observations constitute the central findings of this study and are discussed in detail later. The topological feature of nonlinear transport in TaAs becomes apparent upon characterizing the temperature dependence of nonreciprocal coefficient $\gamma'(T)$ is shown in Fig. 2d. The $\gamma'$ typically arises from the relativistic correction to the band structure[33] and is generally difficult to detect even at low temperatures, let alone at room temperature. Nevertheless, the topological protection of Fermi arcs renders them resilient to thermal fluctuations, enabling the observation of nonlinear transport in TaAs even at room temperature (inset of Fig. 2d). We observed that the sign of $R^{2\omega}(T)$ remains unchanged irrespective of the Hall coefficient $R_H$(T), ruling out a thermoelectric origin, as reported in NaAs nanobelts[34] , which the sign would otherwise follow the dominant carrier type indicated by $R_H$(T) (Extended Data Fig.4).

To verify the topological origin, we compared measurements on the topologically trivial left and right surfaces (Extended Data Fig.5), where Fermi arcs are absent due to mirror symmetry in the *ab*-plane. The normalized $R^{2\omega}(B)$ signals from these trivial surfaces are an order of magnitude smaller than those from topological surfaces, with no sign-reversal features observed in any device measured. When the magnetic field is tilted toward the *c*-axis, pronounced SdH oscillations appear in $R^{2\omega}(B)$, with FFT frequencies matching Fermi pockets identified from linear transport, indicating bulk-dominated signals on trivial surfaces. This stark contrast confirms that the pronounced and sign-reversed nonlinear transport is unique to the topological surfaces of TaAs.

**Nontrivial thickness dependence**

The emergence of topological surface states is often verified through the thickness dependence of electrical transport[35], where thickness refers to the distance between opposing topological surfaces. To probe this, we fabricated a series of FIB-milled devices with thicknesses ranging from 10 μm down to 0.45 μm (see Extended Data Fig.1). Figure 3 displays the thickness dependence of the linear first-harmonic resistivity $\rho^{1\omega}$ (top panel) and the $R^{2\omega}$ (bottom panel). It is well known that FIB milling creates an amorphous surface layer typically 10-20 nm thick[14], inevitably introducing surface damage. In thicker devices, the impact of this damage is negligible for the bulk transport, as indicated by the nearly constant $\rho^{1\omega}$ at 5 K. However, as the thickness decreases below 1 μm, the influence of surface disorder becomes more pronounced, leading to a noticeable increase in residual linear resistivity, as marked by the arrow in the top panel of Fig. 3.

In contrast, as shown in the bottom panel, the corresponding nonlinear second-harmonic sheet resistance $R_s^{2\omega}$ remained nearly constant across all the measured thickness down to 0.45 μm, highlighting the robustness of the surface conduction channel. This behavior aligns with the topological protection of surface states in Weyl semimetals, where spin-momentum locking suppresses backscattering[36], rendering the nonlinear transport resilient to disorder. The nearly thickness-independent $R_s^{2\omega}$ thus naturally reflects the topologically protected nature of Fermi arcs. Further support for this surface origin comes from the thickness dependence of the nonreciprocal coefficient, which becomes consistent and sample-independent only when normalized using a two-dimensional, rather than three-dimensional, model (Extended Data Fig.6). Recently, topologically protected surface conduction channels were observed in NbP thin films, where the linear resistivity $\rho^{1\omega}$ was found to decrease as the thickness was reduced below 80 nm[37].

## Higher-order nonreciprocal components

In addition to the sign-reversed $R^{2\omega}$ feature, a detailed analysis of the angular dependence $R^{2\omega}(\theta)$ reveals anomalous higher-order components, providing more definitive evidence for Fermi-arc-induced nonlinear transport. Specifically, the emergence of nonlinear nonreciprocal transport requires the simultaneous breaking of $\mathcal{T}$ symmetry and a mirror symmetry perpendicular to the transport direction. When subjected to an external magnetic field, Weyl nodes, as monopoles of Berry curvature, shift in opposite directions according to their fixed chirality. As illustrated in Fig. 4a, a magnetic field applied along the $B_x$ direction significantly breaks the mirror symmetry $M_y$. The broken $M_y$ symmetry permits nonlinear transport exclusively along its normal $y$ direction, accounting for the geometric origin of $B\sin\theta$ term in $R^{2\omega}(\theta)$. A more general geometric analysis yields the same $\sin\theta$ dependence when the magnetic field and current are applied in arbitrary in-plane directions (Supplementary information). Further constrained by $\mathcal{T}$ symmetry breaking, only the odd powers of $(B\sin\theta)^N$ with N being an odd integer are allowed. Therefore, the nonlinear transport is expressed as

$$\begin{aligned} R^{2\omega}(I,B) &= R_0 I(\gamma_{1st} B\sin\theta + \gamma_{3rd} B^3 \sin^3\theta + \cdots + \gamma_{Nth} B^N \sin^N\theta) \\ &= R_0 I(\gamma_{1st} B + \frac{3}{4}\gamma_{3rd} B^3)\sin\theta - \frac{1}{4}\gamma_{3rd} B^3 \sin 3\theta + \cdots) \end{aligned} \quad (1)$$

and the right equation is terminated at the third term $N = 3$ for clarity. These symmetry considerations indicate that $R^{2\omega}(\theta)$ contains a prominent $\sin 3\theta$ component when higher-order field terms become non-negligible. Figures 4b and 4c show polar plots of the measured $R^{2\omega}(\theta)$ on the top and bottom surfaces at 8.9 T, clearly resolving a $\sin 3\theta$ pattern with reversed sign. The $\sin 3\theta$ component is absent below 1 T, aligning with the field threshold where the $R^{2\omega}(B)$ deviates from linearity (see Extended Data Fig.8).

Notably, a similar $\sin 3\theta$ pattern in $R^{2\omega}(\theta)$ has been reported in twisted trilayer graphene[38], where it originates from the $C_3$ rotational symmetry encoded in the valley degree of freedom. In contrast, TaAs lacks $C_3$ symmetry, and higher-order terms such as $\sin 5\theta$ are resolved in other devices, such as D1 (see Extended Data Fig.9), unambiguously validating the higher-order origin of the $\sin 3\theta$ component. When examining the reproducibility of higher-order components across different TaAs devices, we observe two distinct types of $R^{2\omega}(\theta)$ patterns occurring with approximately equal probability (Extended Data Figs. 7-10). This discrepancy is attributed to the presence of two distinct Fermi arc topologies in the same Weyl semimetal, as experimentally observed by angle-resolved photoemission spectroscopy[39-41].

Since the above symmetry arguments are not material-specific, such higher-order terms are generally allowed. This naturally raises the question of why these higher-order components in nonlinear transport, clearly present in our raw data, have not been reported previously. To theoretically explore the Fermi-arc-induced nonlinear response, we considered the Boltzmann transport of the Weyl-semimetal surface state under the Zeeman effect, given that the magnetic field is parallel to the surface. Although nonlinear contributions represented in Eq. (1) are found from the energy shift and elongation of Fermi arc, reproducing the observed signals at the low threshold field (~1 T) in TaAs requires an unphysically large $g$-factor (>1000), indicating inadequacy of the simple Boltzmann treatment of band-dispersive nonreciprocity. A similar issue arises in the theoretical analyses of Rashba system, where an excessively large $g$-factor is also required to produce a third-order $B^3$ term at low fields[30]. These results highlight the limitations of the semiclassical framework and suggest that the observed higher-order terms cannot be fully accounted for by a simple Zeeman effect or Rashba-type spin-orbit coupling. In our present treatment, wavefunction effects are neglected as the band dispersion is often deemed the leading contribution. However, Fermi-arc endpoints are singular points where bulk and surface states merge[16] and divergent quantum geometric quantities[15] are theoretically predicted. This singularity, which has not been considered in the above calculations, may underlie the anomalously large $B^3$ term observed in the present experiment. In this regard, a promising direction for the future, indicated by the present result, involves careful but informative investigation of relaxation time dependence and identification of various new contributing mechanisms[42-44].

**Concluding remarks**

Our results demonstrate that nonlinear transport enables direct access to surface contributions even when bulk transport dominates the linear response. Detecting topological surface states in semimetals via charge transport remains challenging due to dominant bulk contributions, unlike in topological insulators. Established methods such as Weyl orbit[7] and quantum Hall measurements[8] rely on linear response and ultrathin devices, which are difficult to achieve in materials like TaAs. The sign-reversed and higher-order nonlinear transport signatures revealed here serve as clear evidence of Fermi-arc contributions, providing a novel and complementary approach to probing topological surface states. Since over 15% of natural materials are predicted to be topological semimetals[17-19], these findings open new avenues for nonlinear transport studies and their broad application.

Topologically protected surface states exhibit intrinsic robustness against disorder and thermal fluctuations. In topological insulators such as magnetic $(Bi,Sb)_2Te_3$, nonlinear transport arising from chiral edge states requires carrier doping to generate a finite resistive response[32]. By comparison, gapless Weyl nodes in Weyl semimetals facilitate quasiparticle scattering between surface and bulk states, enabling dissipative nonlinear transport without doping. This realization of robust nonlinear responses positions Weyl semimetals as promising platforms for surface-switchable topological rectification and room-temperature nonlinear electronics.

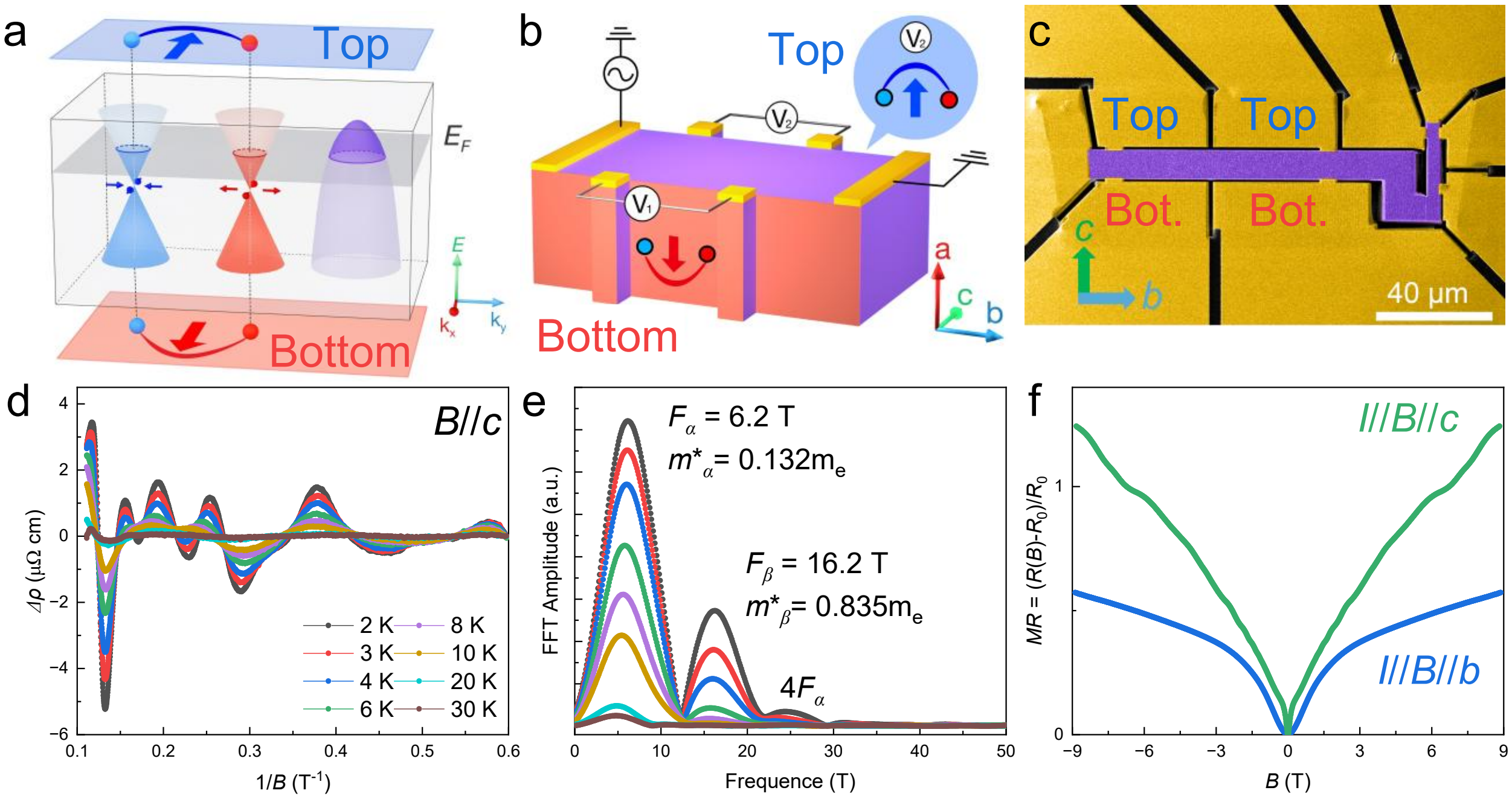

**Fig. 1 | Chirality in TaAs and limitations of linear transport. a,** Schematic illustration of two kinds of chirality in the Weyl semimetal TaAs. In the bulk, Weyl fermions possess opposite chirality, indicated by red and blue arrows. On the top and bottom surfaces (normal to *c*-axis), the associated Fermi arcs exhibit reversed velocity as marked by arrows. TaAs hosts both electron-like Weyl fermions and topologically trivial hole pockets (purple). For clarity, only one representative pair of the total 12 Weyl node pairs is shown. **b,** Schematic of focused ion beam (FIB) fabricated device and measurement configuration. The top and bottom surfaces are exposed at the sidewalls. **c,** Scanning electron microscope image of TaAs device D1. The TaAs microstructures and electrodes are highlighted in purple and yellow, respectively. **d,** Shubnikov–de Haas (SdH) quantum oscillations as a function of inverse magnetic field from 2 K to 30 K in TaAs device D1. A polynomial background has been subtracted from the resistivity. **e,** Corresponding fast Fourier transform (FFT) spectrum of SdH quantum oscillation. Two frequency components, $F_\alpha = 6.2$ T and $F_\beta = 16.2$ T, are clearly resolved, while no signature of the Weyl orbit is observed in this thick device. **f,** Chiral anomaly experiments along the *c*- and *b*-axes at $T = 5$ K. Due to the high-mobility of electrons ($\mu_e > 10^5$ cm$^2$V$^{-1}$s$^{-1}$) and non-Weyl background contributions from hole-type carriers, the chiral anomaly signature is masked in TaAs up to 9 T.

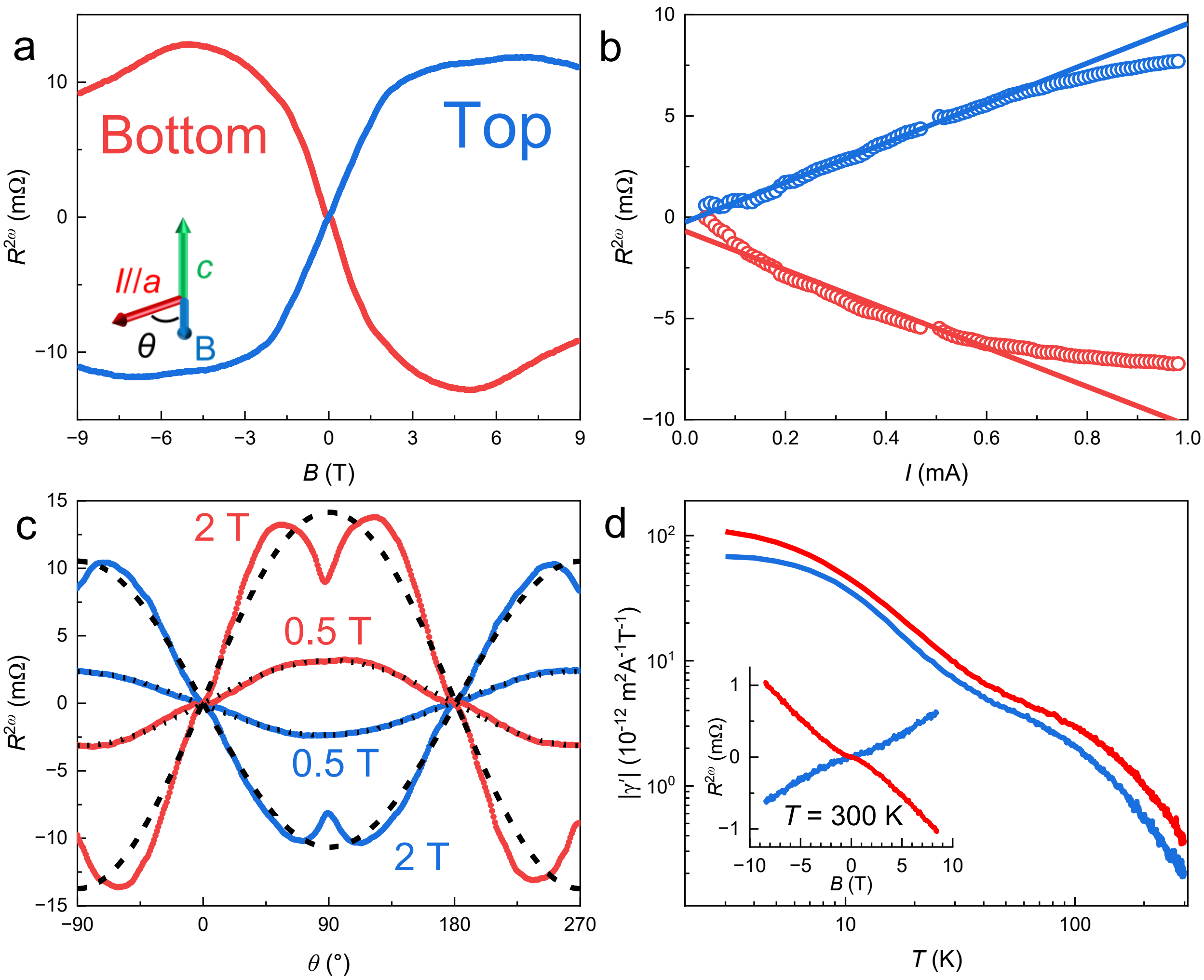

**Fig. 2 | Surface-switchable nonlinear transport on opposing topological surfaces in TaAs. a,** Field dependence of second-harmonic resistance $R^{2\omega}$ at $T = 5$ K, measured with the angle $\theta$ set to -90°. Data from the top and bottom topological surfaces are shown in blue and red, respectively. The inset illustrates the angle $\theta$ between the current and magnetic field. **b,** $R^{2\omega}$ as a function of current $I$ at 5 K, 1T. **c,** Angular dependence of $R^{2\omega}$ measured under 5 K. Additional features become evident under higher magnetic field (e.g. 2 T). The black dashed lines represent the sine wave fitting. **d,** Temperature dependence of the nonreciprocal coefficient $\gamma' = 2R^{2\omega}/R_0JB$. $R_0$ is the zero field linear resistance and $J$ is the current density. The inset displays the pronounced nonlinear transport at room temperature. All data were obtained from TaAs device D6.

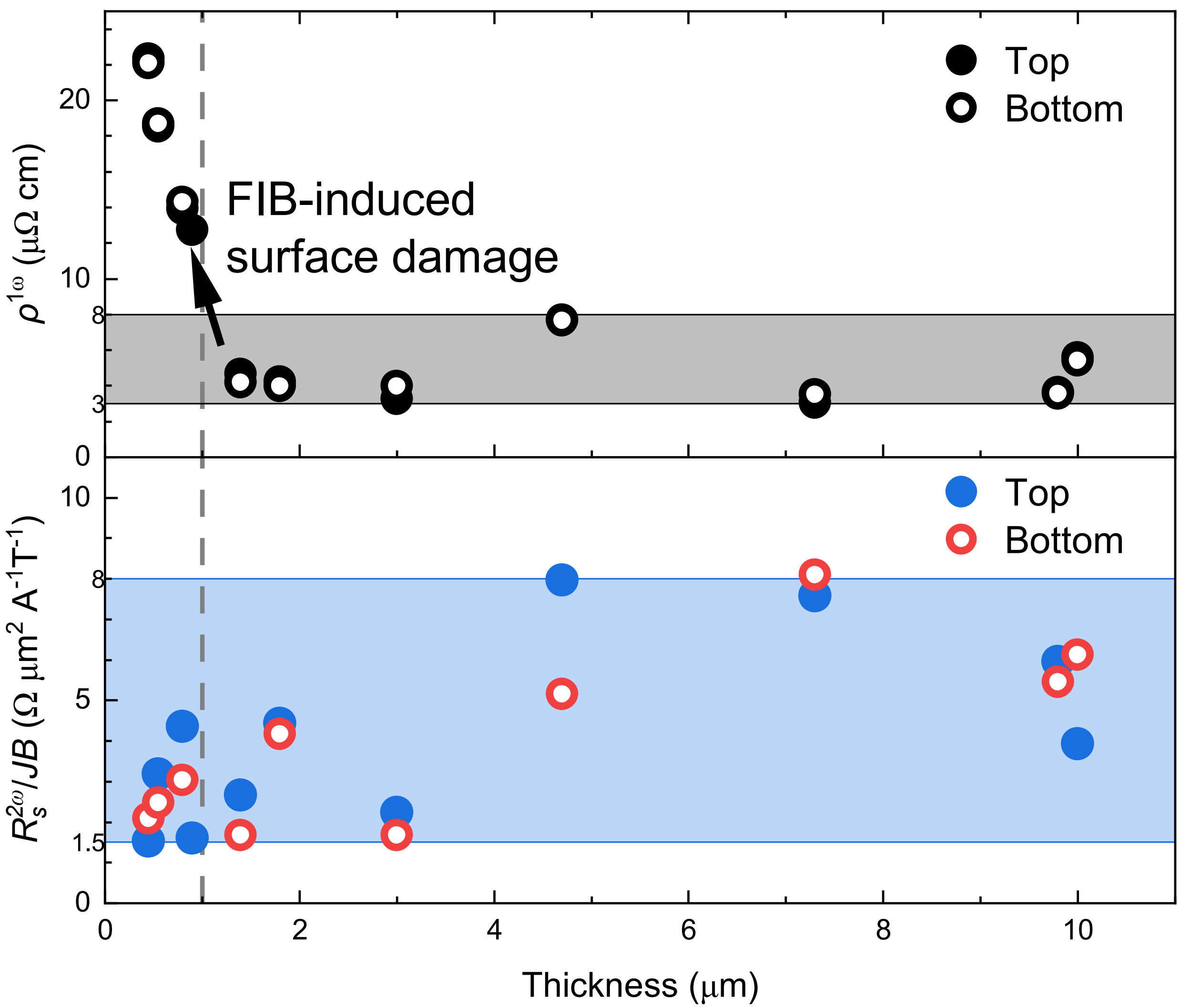

**Fig.** 3 | **Nontrivial thickness dependence of linear and nonlinear resistivity.** The second-harmonic sheet resistance $R_s^{2\omega}$, normalized by current density $J$ and magnetic field $B$, is shown for multiple devices. Solid and open circles represent data from the top and bottom topological surfaces, respectively.

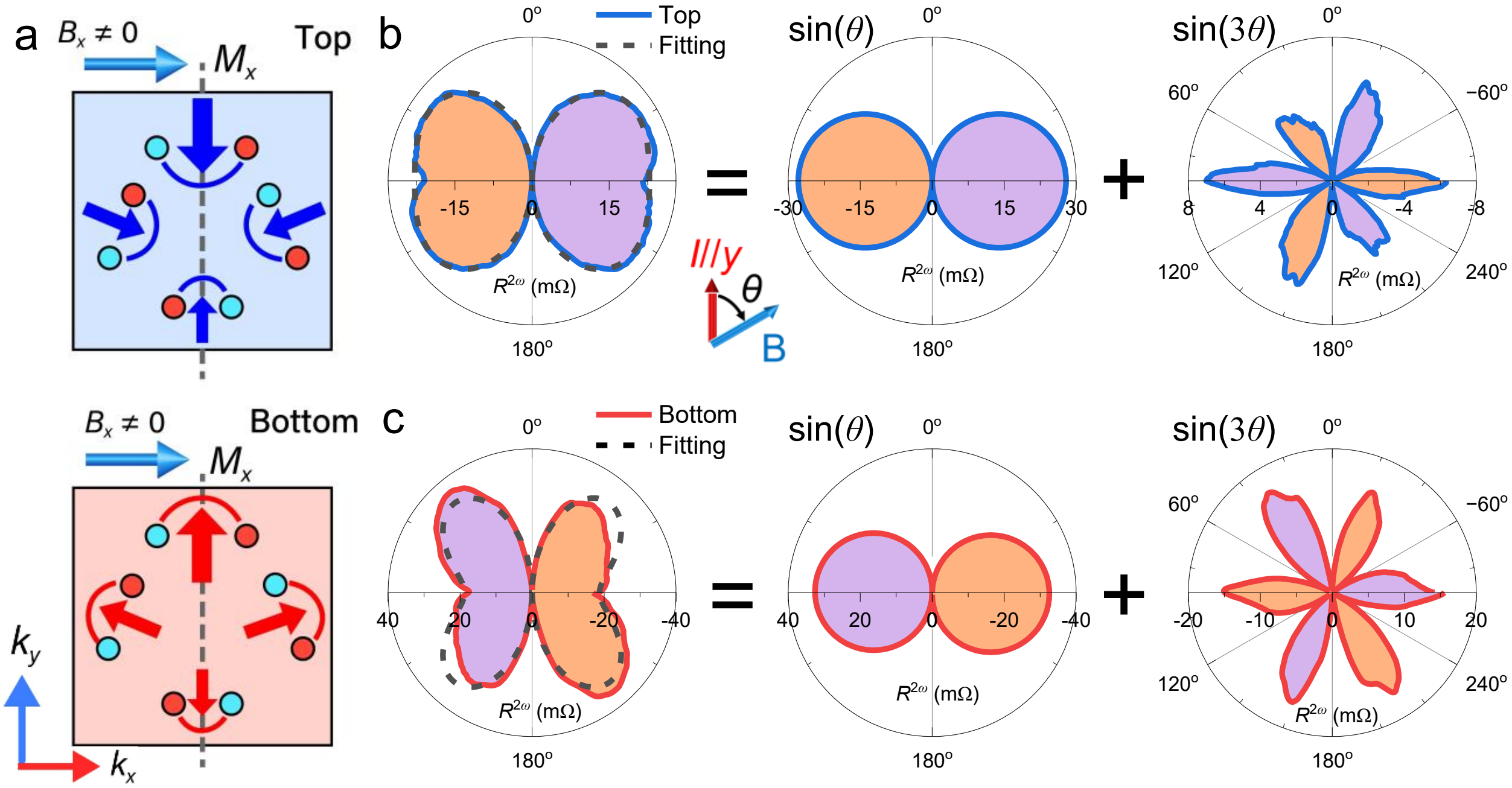

**Fig. 4 | Third-order angular components in Fermi-arc-induced nonlinear transport. a,** Illustration of magnetic field $B_x$ induced $M_y$ mirror symmetry breaking. TaAs hosts 12 Weyl node pairs in total, of which only four are depicted here for clarity. The displacement of the Weyl nodes is identical on the top and bottom surfaces, but the opposite directionality of their Fermi arcs results in a sign-reversed response. **b,** Polar plot of angular-dependent second-harmonic resistance $R^{2\omega}(\theta)$. The blue raw data is well-fitted by the black dashed curve, which is the sum of $\sin(\theta)$ and $\sin(3\theta)$ components. The right panel shows the fitted $\sin(\theta)$ component and the residuals, highlighting a clear three-fold symmetry. Purple and orange denote positive and negative values of $R^{2\omega}$, respectively. **c,** Same as b, but for the bottom surface, displayed in red.

# Methods

## Devices fabrications

We employed state-of-the-art methods to fabricate TaAs devices using the focused ion beam (FIB) technique[14], which prevents extrinsic nonreciprocal signals introduced by FIB-assisted platinum deposition[29] and significantly improves electrode contact quality. Extended Data Fig. presents all the TaAs devices studied. The entire fabrication process can be summarized in four steps.

**First**, *preparing the lamella from bulk samples.* The crystallographic orientation of TaAs was first determined using Laue diffraction, and the samples were mounted accordingly. A lamella with typical dimensions of $150 \times 60 \times 3$ μm³ was then cut from the crystal. In this step, we used a Hitachi NB5000 FIB operating at 40 kV. A high current of 69 nA was applied to dig the grooves, followed by a lower current of 3.7 nA under grazing incidence (±1.5°) to polish the sidewalls into a flat shape.

**Second**, *transferring the lamella to the substrate.* A droplet of Araldite epoxy glue (approximately 200 μm in diameter) was placed onto a sapphire substrate pre-patterned with gold contact pads. Using an ex-situ micromanipulator, the lamella was vertically lifted from the bulk crystal and horizontally placed onto the glue droplet. Capillary forces from the liquid epoxy ensured continuous and smooth contact between the lamella and the substrate.

**Third**, *evaporating gold electrical contacts.* This step is crucial for reducing contact resistance. The epoxy droplet was first baked at 150 °C for one hour to remove any trapped air bubbles. Afterward, the lamella surface was briefly etched with argon to improve adhesion. Using a homemade shadow mask, Ti/Au layers (5/200 nm) were deposited by electron-beam evaporation to cover both the lamella and the contact pads. These metal layers serve as electrical contacts with a typical contact resistance of only a few ohms.

**Finally**, *shaping the measurement region.* The device was returned to the FIB chamber for the final fabrication step. A FEI Helios 5 UX FIB system was used to offer better operational flexibility. The gold layer covering the device was carefully removed using FIB at 1.6 nA and 5 kV. The overall device shape was then defined using FIB milling at 0.79 nA and 30 kV. To avoid potential artifacts from asymmetric processing, the top and bottom regions were simultaneously cut under identical conditions. Finally, the outer gold layer was selectively removed to ensure that current flow during transport measurements was confined to the intended region.

## Electrical transport measurements

The electrical transport measurements were performed using a Quantum Design PPMS-9 system over a temperature range from 2 to 300 K. During nonlinear transport measurements, both the first-harmonic resistance $R^{\omega}$ and second-harmonic resistance $R^{2\omega}$ (with a -90° phase offset) were recorded using Stanford SR830 lock-in amplifiers with an input frequency of 137.77 Hz in the nonlinear transport measurements. The applied current was carefully adjusted case by case to minimize Joule heating effects.

## Quantum oscillations analysis

Shubnikov–de Haas (SdH) quantum oscillations were routinely measured to verify the pristine properties of FIB-fabricated devices. We examined SdH oscillations across multiple TaAs devices and observed consistent results. Extended Data Fig.2 presents representative SdH data for device D1, measured from 2 K to 30 K. The corresponding background-subtracted oscillatory component and its fast Fourier transform (FFT) spectrum are shown in the main text. The temperature dependence of FFT amplitude is fitted well by the Lifshitz–Kosevich formalism $A(T) = A_0 2\pi^2 k_B T / (\hbar\omega_c) / \sinh(2\pi^2 k_B T / \hbar\omega_c)$. Here $k_B$ is the Boltzmann constant and $\omega_c$ = eB/$m^*$ is the cyclotron frequency. The fitted effective mass of $m^*$ are 0.057 $m_e$ for $F_\alpha$ and 0.139 $m_e$ for $F_\beta$, respectively, consistent with a previous report[26].

**Nonlinear transport in trivial surfaces**

Due to in-plane mirror symmetry, Fermi arcs appear exclusively on the top and bottom surfaces. As a comparative study, we also measured the left and right (side) surfaces of TaAs devices (Extended Data Fig. 5), which are topologically trivial. Three distinct features were observed:

1. When the magnetic field is aligned along the *c*-axis, clear SdH quantum oscillations appear in the $R^{2\omega}$(B) data, with FFT frequencies corresponding to Fermi pockets resolved in the linear response regime.
2. The normalized $R^{2\omega}$(B) signals from the tside surfaces are significantly smaller than those from the topological surfaces, and no clear sign-reversal features are observed.
3. While the angular dependence $R^{2\omega}$(B) fits well to a simple sin($\theta$) function at low magnetic fields, it becomes completely irregular at higher fields above 1 T. These disordered patterns may arise from the interference of SdH oscillations and are clearly distinct from the higher-order modulations observed on the topological surfaces.

Taken together, these findings support the conclusion presented in the main text that the unique nonlinear transport observed on the top and bottom surfaces originates from topological Fermi arcs.

**Thickness dependence**

Technically, the $Ga^{2+}$ ion beams are accelerated at 30 kV, and their energy is fully absorbed within an effective depth of 10-20 nm from the surface, rendering the milled surface amorphous and introducing inevitable surface damage in FIB-fabricated devices[14]. In the case of TaAs devices, FIB milling induces amorphous Ta-rich surface layers, leading to artificial surface superconductivity[45]. Figure S4 shows the temperature-dependent first-harmonic resistivity of TaAs devices with varying thicknesses. The superconducting transition consistently emerges below 3 K across all devices. This observation is consistent with previous reports of FIB-induced superconductivity in Weyl semimetal TaAs[45]. The existence of FIB-induced amorphous layer is directly confirmed by transmission electron microscopy on FIB devices fabricated under the same processing conditions[46].

Extended Data Fig. 6 shows the thickness dependent nonreciprocal coefficient $\gamma'$, analyzed using both three-dimensional (3D) and two-dimensional (2D) models. The reduced nonreciprocal coefficient, defined as $\gamma'=\gamma A$, is normalized by the sample cross-sectional area $A$ and is therefore expected to be independent of thickness. We first examined the thickness dependence of the 3D $\gamma'$, as is typically done for bulk-induced nonlinear responses. However, the 3D $\gamma'$ values increase monotonically with sample thickness, yielding an unphysical result. This discrepancy arises because the linear (first-harmonic) resistivity is dominated by bulk transport, while the nonlinear (second-harmonic) response originates primarily from the surface.

To resolve this issue, we instead analyze the ratio between the second-harmonic sheet resistance $R_s^{2\omega}$ and the first-harmonic resistivity $\rho^{1\omega}$, from which the 2D γ′ is extracted. The thickness dependence of the 2D γ′ remains nearly constant for devices with thickness above 1 μm but gradually decreases with thickness for devices thinner than 1 μm due to an overestimation of $\rho^{1\omega}$ caused by FIB-induced surface damage. The stark contrast between the thickness dependence of the 3D and 2D γ′ further confirms the surface-originated nature of the nonlinear transport.

**Angle-resolved R2ω(θ) and higher-order components**

We observed two distinct types of $R^{2\omega}(\theta)$ patterns across different TaAs devices, which we designate as type-A and type-B. Although visually different, both can be consistently described by higher-order modulations (Extended Data Fig. 7), with the $\sin(3\theta)$ component exhibiting opposite signs between the two types. As discussed in the main text, these patterns correspond to two distinct Fermi surface topologies of the Fermi arcs in Weyl semimetals.

Extended Data Figs. 8 and 9 present the detailed magnetic-field dependence of the type-A and type-B patterns, respectively. At low fields, the $\sin(3\theta)$ component is negligible, but it becomes increasingly pronounced at higher fields, in line with its higher-order character. A more detailed scrutiny of angular dependence reveals that up to the $\sin(3\theta)$ component can be resolved in type-A, whereas type-B exhibits higher-order terms up to sin(7θ) (figs. S6 and S7). In real devices, nonlinear transport reflects the coexistence of both domain types, leading to a more complex $R^{2\omega}(B)$ response than the simple sign-reversal expected for a single domain. Consequently, the sign of $R^{2\omega}(B)$ depends on the relative domain composition: when dominated by the same domain type on both top and bottom surfaces, the sign is fully reversed; when type-A and type-B coexist in comparable proportions, higher-order terms of opposite sign can cause internal sign reversals within a single surface (Fig. S8).

In contrast, the $\sin(3\theta)$ components are highly reproducible across all measured TaAs devices (Extended Data Fig. 10). Statistically, type-A and type-B domains occur with roughly equal probability, consistent with a random domain distribution. While a full theoretical understanding of the higher-order terms in Fermi arcs remains to be developed, their consistent observation across samples provides strong experimental support. Empirically, these higher-order components serve as clear “smoking-gun” evidence of Fermi-arc-induced nonlinear transport.

**Theoretical model on Fermi-arc-induced higher-order terms**

In this section, we consider the nonlinear response of the surface state of a Weyl semimetal under a magnetic field within the Boltzmann transport. This includes two contributions due to the magnetic field: i) energy shift of surface band/Fermi arc; ii) Fermi-arc elongation due to the expanding surface state region. The detailed derivation is provided in the Supplementary Information. The higher-order term contribution is then obtained as

$$F(B) \propto (0.18 sin\theta)b + (0.09 sin\theta + 0.03 sin3\theta)b^3 + \cdots$$

While this model qualitatively reproduces the higher-order terms, it fails quantitatively. To compare with experimental results, which show a clear third-order contribution above the characteristic field $B^* = 1$T, we adopt a hopping energy scale $t = 0.4$ eV for TaAs[16]. The dimensionless parameter $b$ is obtained as $b = g\mu_B B/t$, where $g$ is $g$-factor and $\mu_B$ is the Bohr magneton. To satisfy the condition $b^3/b = 1$ at $B^* = 1$T for observing third-order term, the required $g$-factor is calculated to be approximately 6900, which is unphysically large.

Similarly, for the Rashba spin-orbit coupling with $\lambda$ = 75 meV [30], the dimensionless parameter is defined as $g\mu_B B/\lambda$. An unphysically large $g$-factor of approximately 1300 is also obtained. These predictions are consistent when considered from the perspective of energy scale. Resolving this discrepancy requires a novel mechanism of nonreciprocal transport that incorporates the divergent quantum geometry at the Fermi-arc endpoints[15].

**Data availability**

The data shown in the figures are available from datadryad.org. Other data that support the findings of this study are available from the corresponding author on request.

**Acknowledgments:** We acknowledge Y. Tokura, Z.X. Shen, M. Kawamura, and C. Zhang for insightful discussions, and X.Z. Yu, Y. Kubota, and M. Murate for providing access to the FIB machines. We thank T. Shitaokoshi and X. Huang for experimental assistance, and Mari Ishida for help in preparing schematic figures. This work was performed at the Nanoscience Joint Laboratory, which are supported by the RIKEN Center for Emergent Matter Science (CEMS), Japan. A portion of the FIB work was supported by the Support Unit for Electron Microscopy Techniques (EMT), Research Resources Division (RRD), RIKEN Center for Brain Science (CBS), Japan. M.H. acknowledge support from Japan Society for the Promotion of Science (JSPS) KAKENHI grant 24H01607, and Japan Science and Technology Agency grant JPMJAP2426, CREST grants JPMJCR1874, PMJCR20T1, JPMJCR20T2, and FOREST grant JPMJFR2238. N.N. acknowledge support from JSPS grants 24H00197, 24H02231, 24K00583, and RIKEN Transformative Research Innovation Platform initiative. X.X.Z. acknowledge the startup funding of Huazhong University of Science and Technology. Y.M.I., Y.M.X., and M.T.B. acknowledge RIKEN Special Postdoctoral Researchers program. ChatGPT was used solely to enhance readability. The manuscript contains no AI-generated content.

**Author contributions:**

D. Li and Y. Iwasa conceived the project. M.H.Z. and Z.A.N. synthesized the crystals. D.L. fabricated the devices with assistance from Y.M.I. Y.F. Y.G. and M.T.B. D.L. performed the transport measurements. Y.M.X. K.J.Z. X.X.Z. and N.N. developed the theoretical framework. D.L. and Y.I. analyzed the experimental results with input from Y.M.I. Y.M.X. I.B. M.H. X.X.Z. T.M. and N.N. D.L. and Y.I. wrote the manuscript with contributions from all authors.

**Competing interests:** Authors declare that they have no competing interests.

**Additional information**

Supplementary information

**Correspondence and requests for materials** should be addressed to Dong Li or Yoshihiro Iwasa.

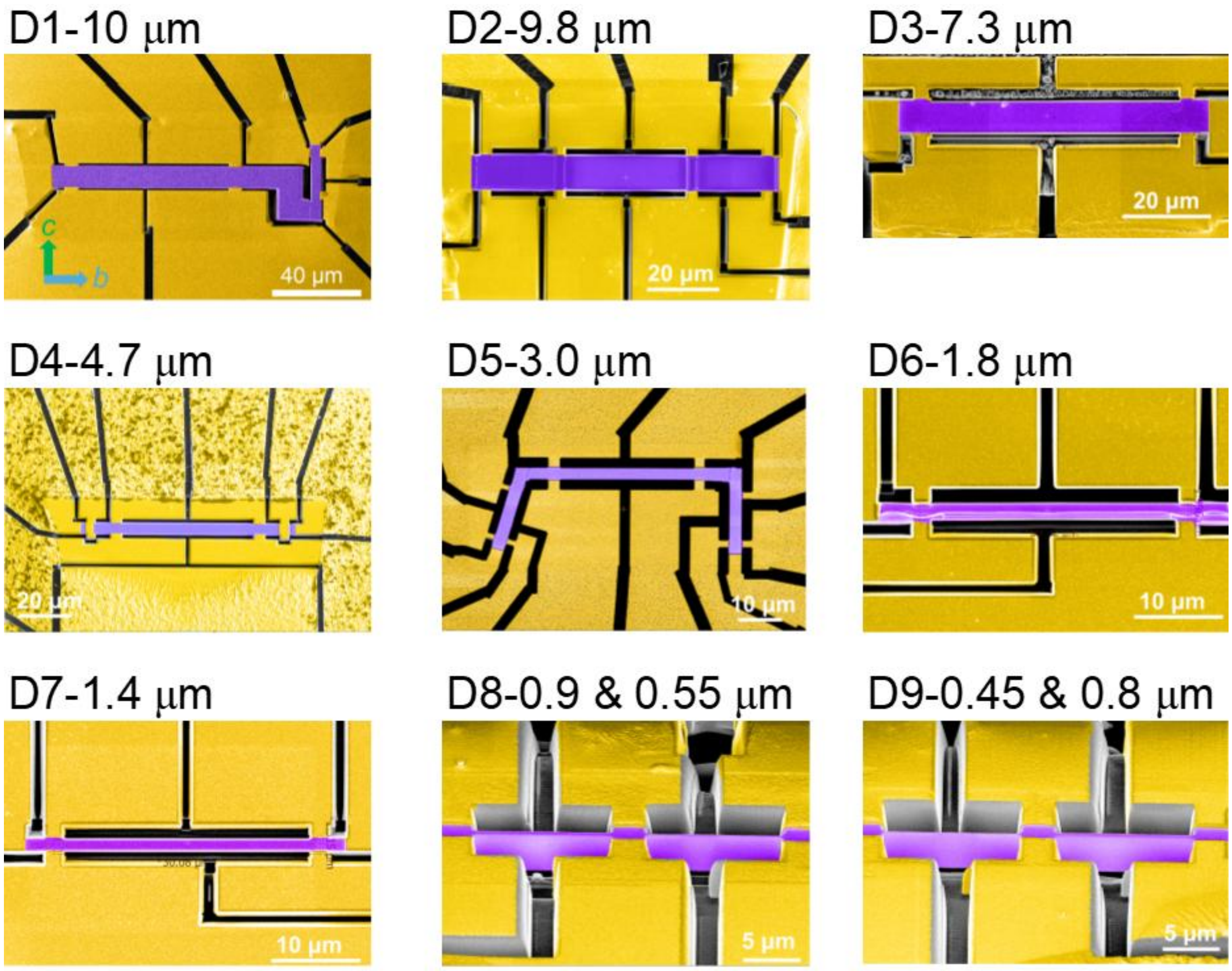

**Extended Data Fig. 1 Scanning electron microscope image of TaAs devices with different thickness**. The thickness, defined as the distance between the top and bottom surfaces, is indicated next to each device symbol. All devices are shown in top view, except for devices D8 and D9, which are imaged at a tilted angle to clarify the sample position.

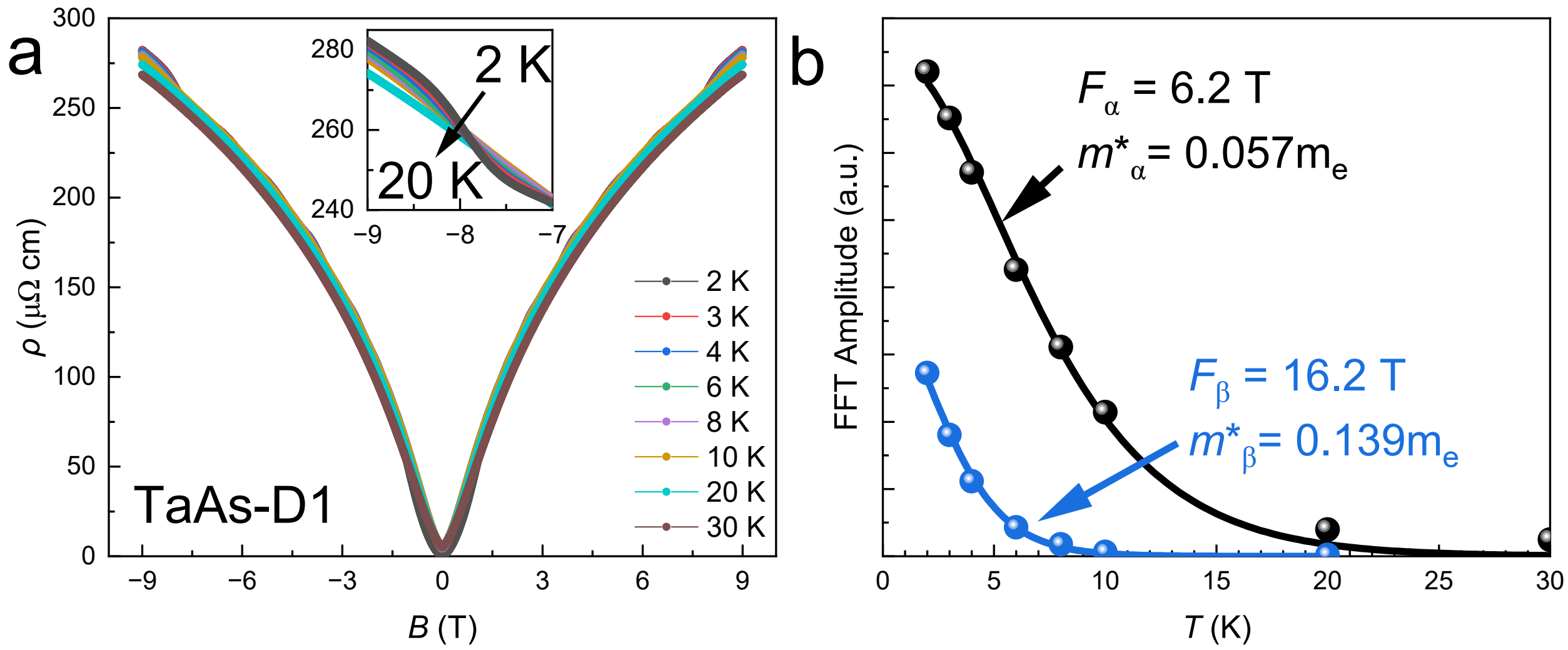

**Extended Data Fig. 2 Shubnikov–de Haas (SdH) quantum oscillations in TaAs device. a,** The field dependence of resistivity in TaAs device D1 from 2 K to 30 K. The inset shows the oscillating behavior from -7 T to -9 T. The magnetic fields were applied along the (001) orientation. **b,** The temperature dependence of the FFT amplitudes following the usual Lifshitz–Kosevich behavior. The fitting results yield the effective electron mass of 0.057 $m_e$ for $F_\alpha$ (black dots) and 0.139 $m_e$ for $F_\beta$ (blue dots), respectively.

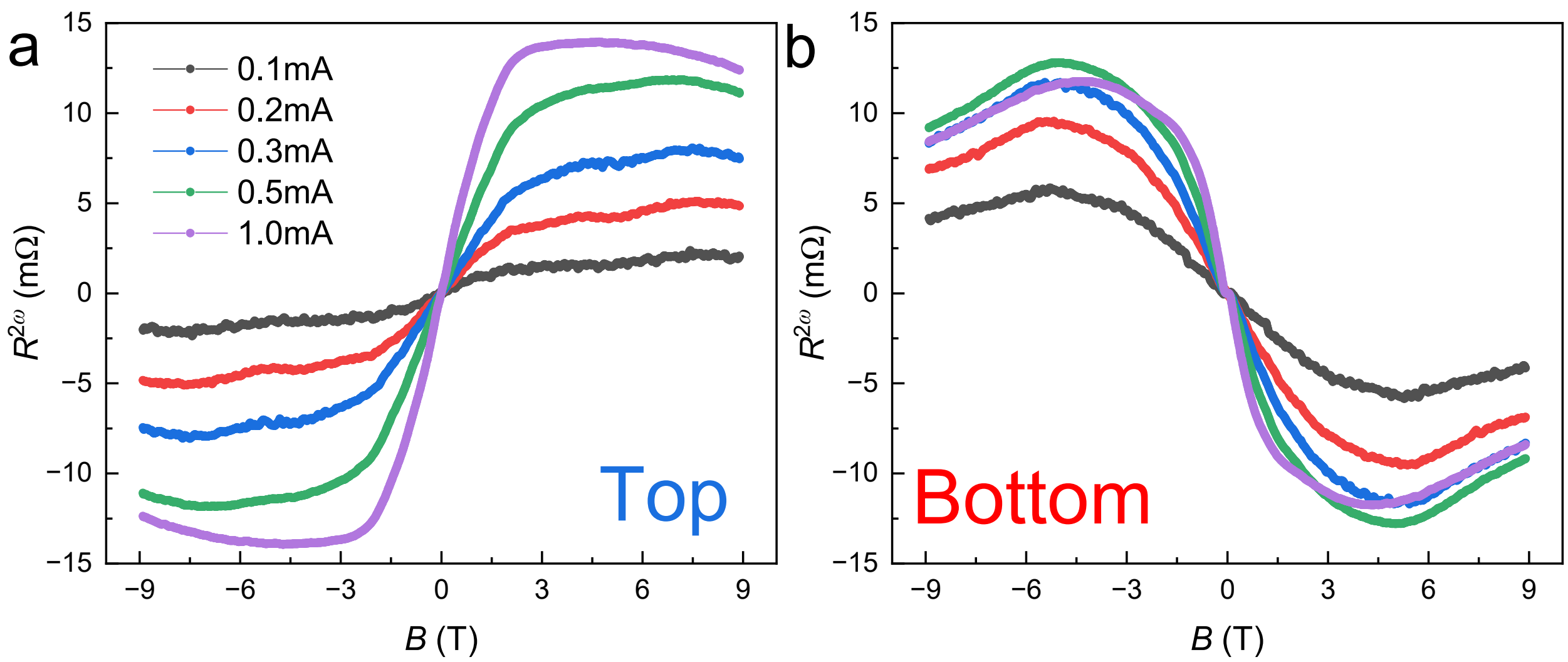

**Extended Data Fig.3 Intrinsic deviation from linearity on the field-dependent second-harmonic resistance $R^{2\omega}$(B). a,** The field-dependent $R^{2\omega}(B)$ measured on the top surface of the TaAs device under various applied currents. The robust deviation behavior are clearly observed across both low and high current regimes, ruling out extrinsic effects such as Joule heating. **b,** same as a, but for the bottom surface.

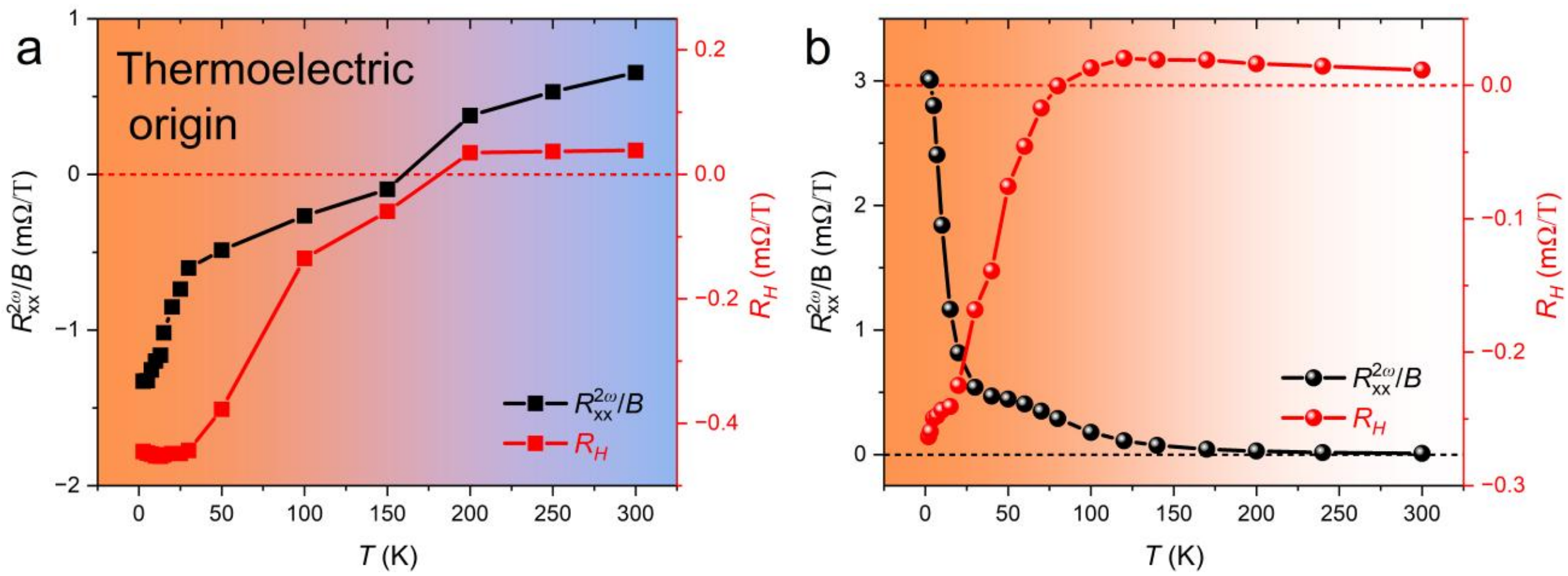

**Extended Data Fig. 4 Exclusion of a thermoelectric origin. a,** Temperature dependence of longitudinal second-harmonic resistance $R^{2\omega}_{xx}/B$ and Hall coefficient $R_H$ in Weyl semimetal NbAs nanobelts. Raw data are extracted from ref[34]. In the case of thermoelectrically induced $R^{2\omega}$, the sign is determined by the dominant carrier type, while the magnitude is less sensitive to temperature. **b,** Comparison of temperature dependence of $R^{2\omega}_{xx}/B$ and $R_H$ in TaAs device D1. The $R^{2\omega}$ data are obtained from the top surface. In contrast, the sign of $R^{2\omega}$ is independent of the Hall sign and diverges at low temperature, thereby directly excluding a thermoelectric origin.

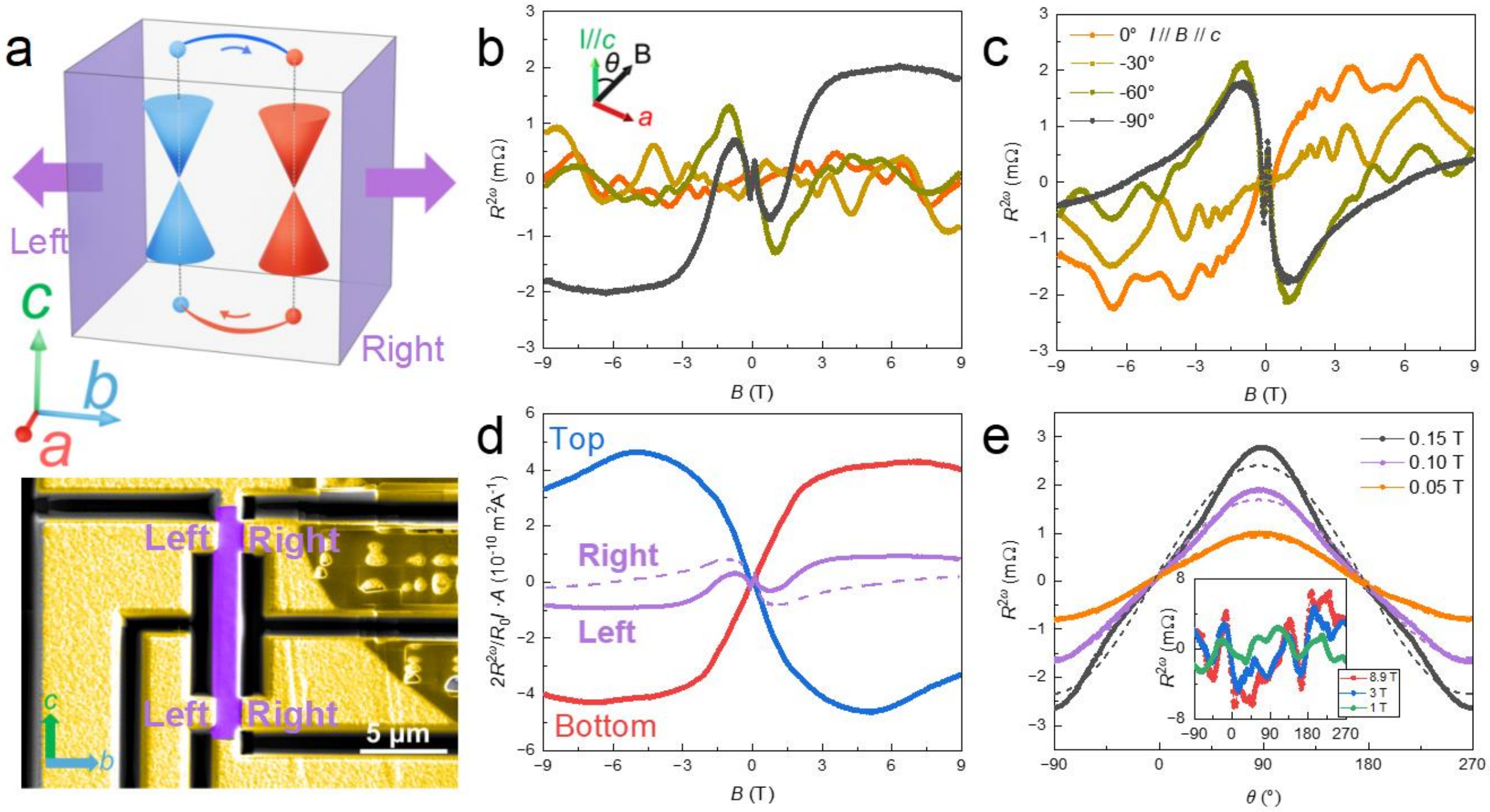

**Extended Data Fig. 5 Nonlinear transport on the topological trivial surfaces in TaAs. a,** Schematic illustration of the measurement configuration on the left and right surfaces (normal to *b*-axis) of the TaAs device, which lack Fermi arcs. The bottom panel shows a scanning electron microscope image of the corresponding device. **b,** Field-dependent second-harmonic resistance $R^{2\omega}(B)$ measured on the left surface of the TaAs device under different angle $\theta$. The inset defines $\theta$ as the angle between magnetic field $B$ and applied $c$-axis current $I$ within the *ac*-plane. **c,** same as (B) but for the right surface. **d,** Comparison of normalized nonreciprocal transport $2R^{2\omega}(B)/R_0J$ between surfaces with (Top and Bottom) and without (Left and Right) Fermi arcs. $R_0$ is the zero field linear resistance and $J$ is the current density. **e,** Angular dependence of $R^{2\omega}$ measured at 5 K under various magnetic fields from left surface. The $R^{2\omega}(\theta)$ curves follow sinusoidal behavior at low fields but are distorted by quantum oscillations at higher fields.

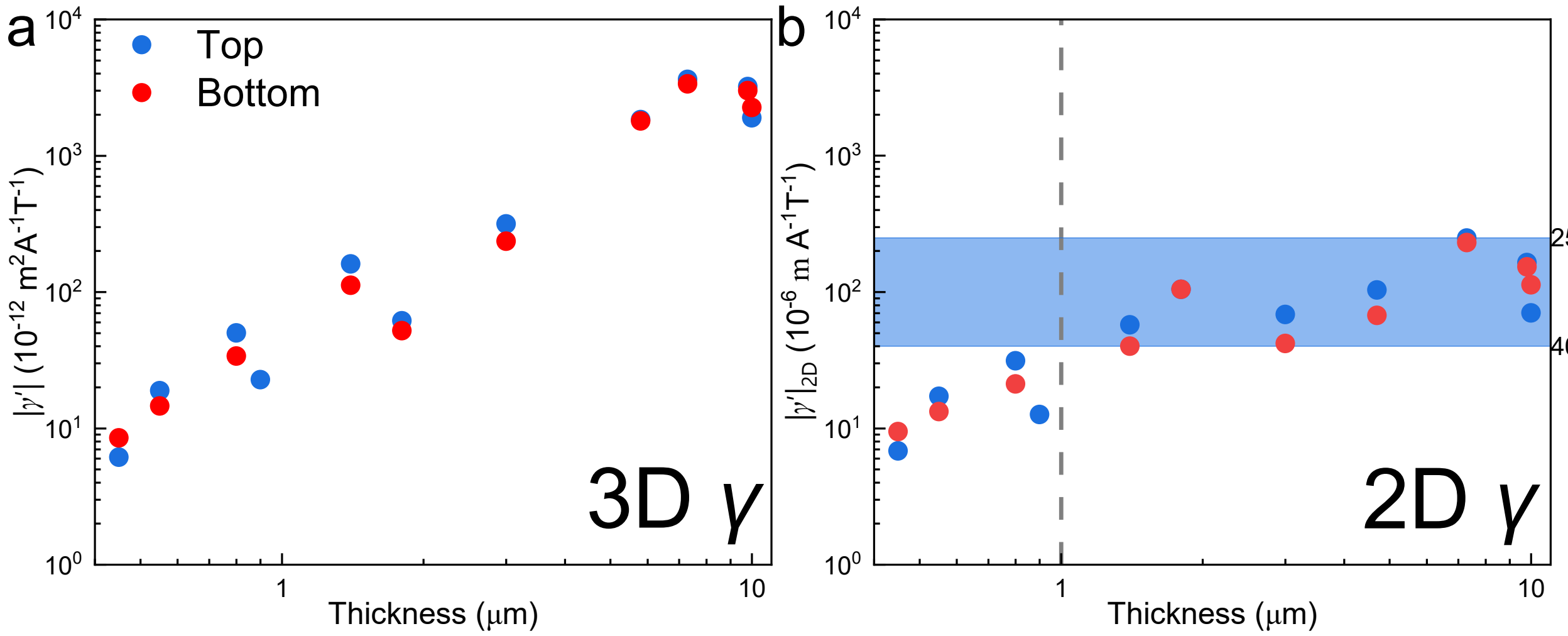

**Extended Data Fig. 6 Thickness dependence of nonreciprocal coefficient. a,** The thickness dependent three-dimensional nonreciprocal coefficient $\gamma' = \frac{2R^{2\varpi}}{R^{1\varpi}JB}$. Absolute values are displayed for clarity. **b,** same as a, but for the two-dimensional nonreciprocal coefficient $\gamma_{2D}' = \frac{2R_S^{2\varpi}}{\rho^{1\varpi}JB}$. The decreas of $\gamma_{2D}'$ occurs in devices with thickness thinner than 1 μm due to an overestimation of $\rho^{1\omega}$ caused by FIB-induced surface damage.

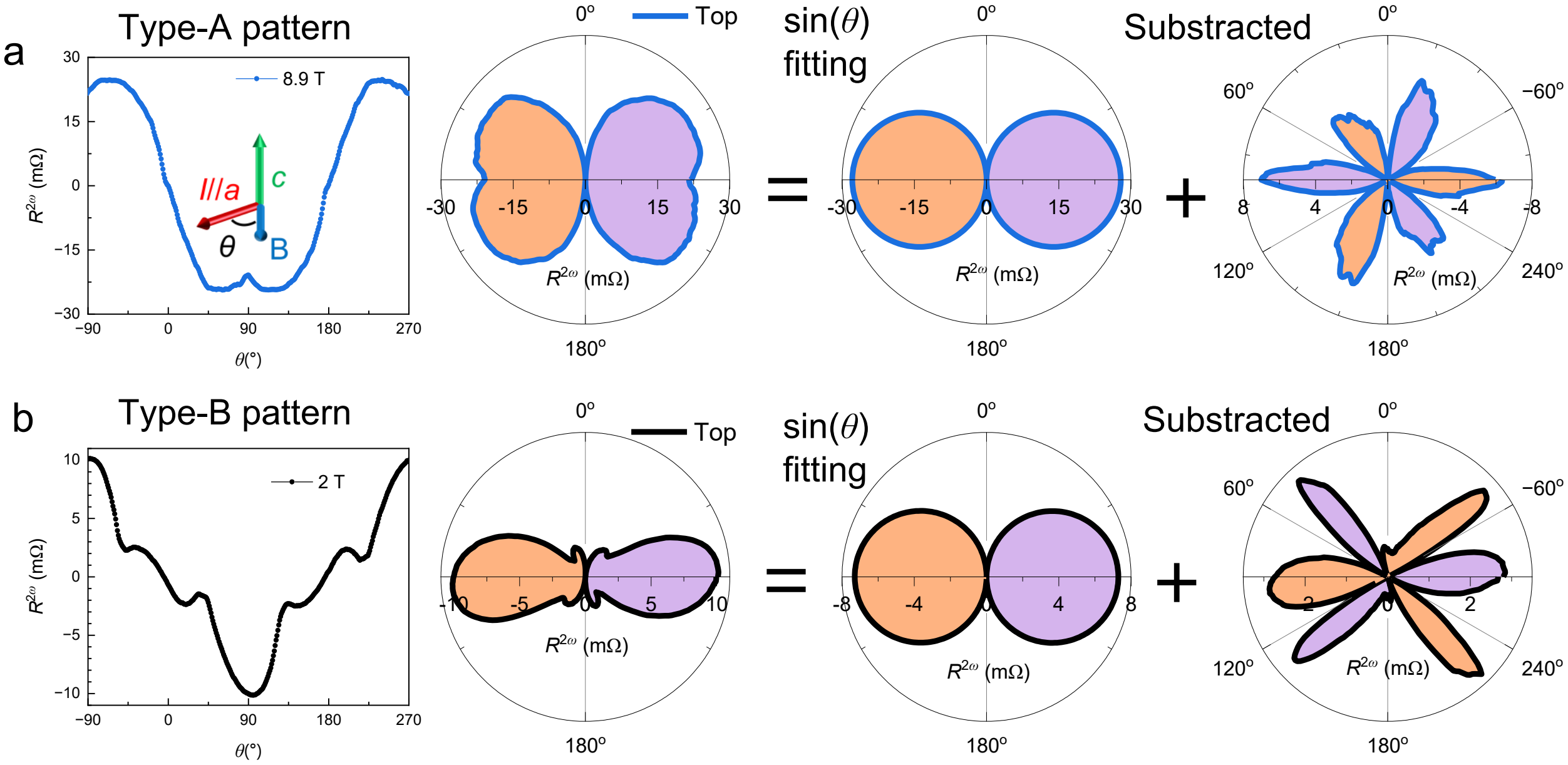

**Extended Data Fig. 7 Two seemingly distinct types of $R^{2\omega}(\theta)$ patterns. a,** Type-A angular dependent second harmonic resistance $R^{2\omega}(\theta)$ measured from top surface of TaAs device D6. Right panels show the corresponding polar plot, showing that the pattern is well described by a combination of $\sin(\theta)$ and $\sin(3\theta)$ components. **b,** same as a, but for the distinct type-B pattern measured from the top surface of TaAs device D1. The sign of the $\sin(3\theta)$ component is reversed between type-A and type-B patterns. The type-B pattern is more strongly affected by higher-order terms such as $\sin(5\theta)$ at high magnetic fields. We present the data at $B$ = 8.9 T for the type-A pattern and at a moderate field of $B$ = 2.0 T for the type-B pattern. Detailed magnetic field dependence is provided in the following Extended Data Figures.

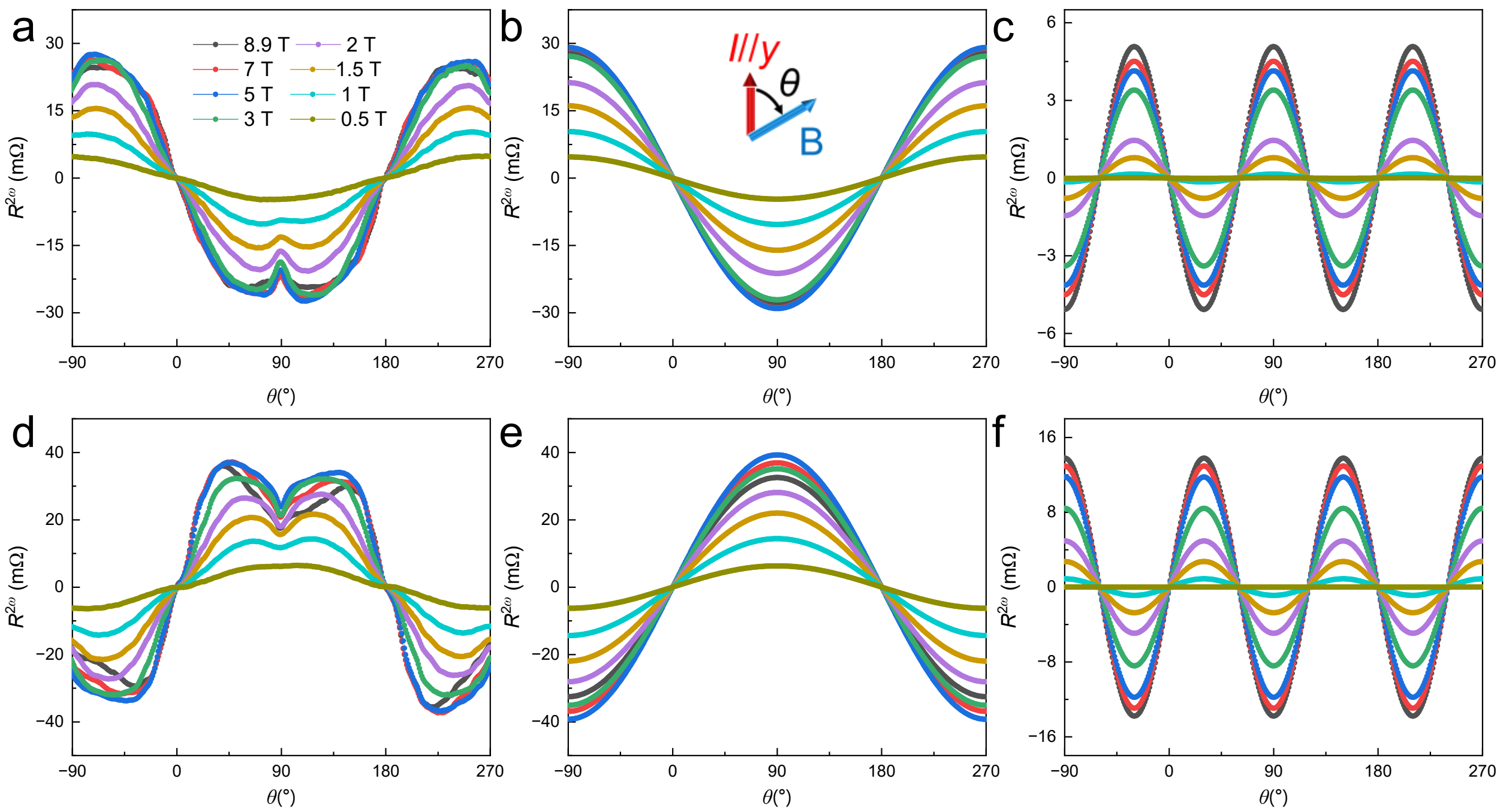

**Extended Data Fig. 8 Field dependence of sin(3$\theta$) modulation in type-A patterns. a,** Raw data of angular dependent second harmonic resistance $R^{2\omega}(\theta)$ measured from top surfaces of TaAs device D6. **b,** The associated fitting curve for the sin($\theta$) component. **c,** The associated fitting curve for the sin(3$\theta$) component. **d-f,** same as a-c, but for the bottom surfaces of TaAs device D6.

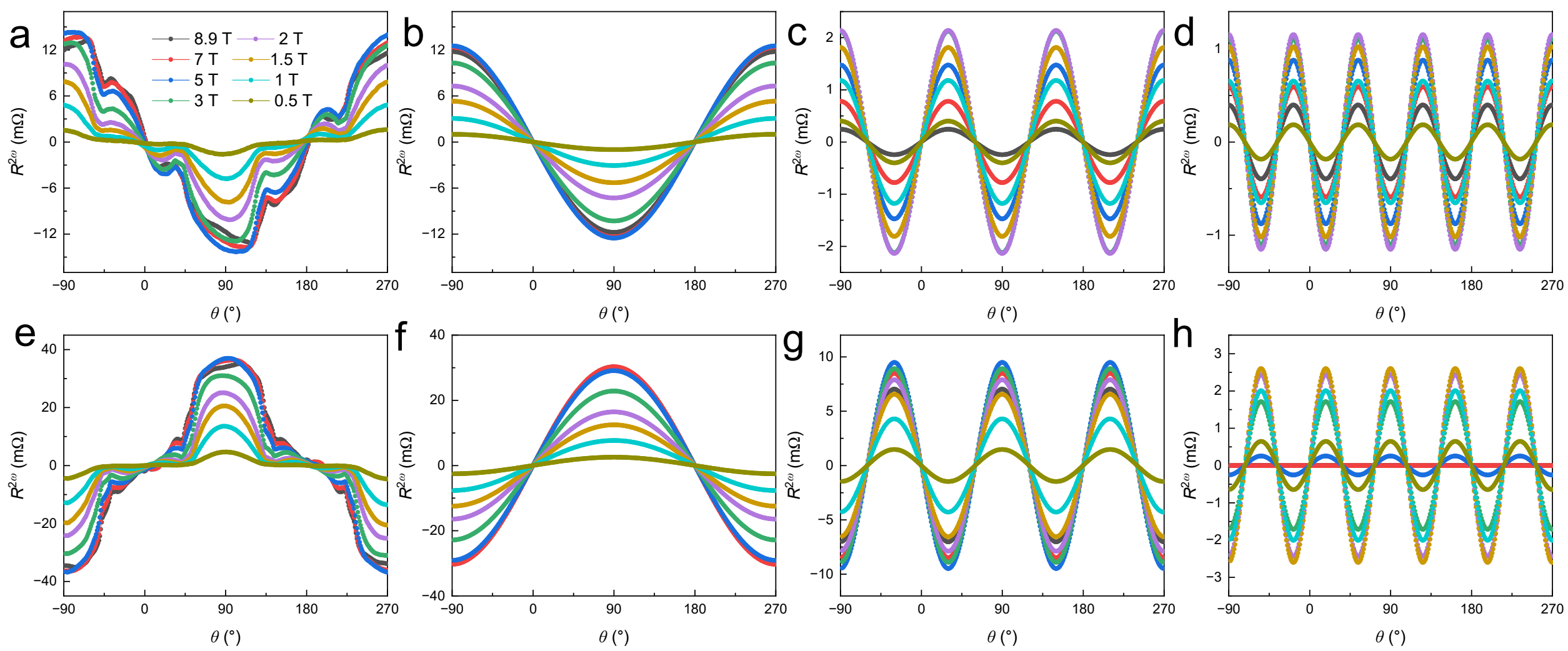

**Extended Data Fig. 9 Field dependence of higher-order modulations in type-B patterns. a,** Raw data of angular dependent second harmonic resistance $R^{2\omega}(\theta)$ measured from top surfaces of TaAs device D1. **b,** The associated fitting curve for the $\sin(\theta)$ component. **c,** The associated fitting curve for the $\sin(3\theta)$ component. **d** The associated fitting curve for the $\sin(5\theta)$ component. **e-h,** same as a-d, but for the bottom surfaces of TaAs device D1.

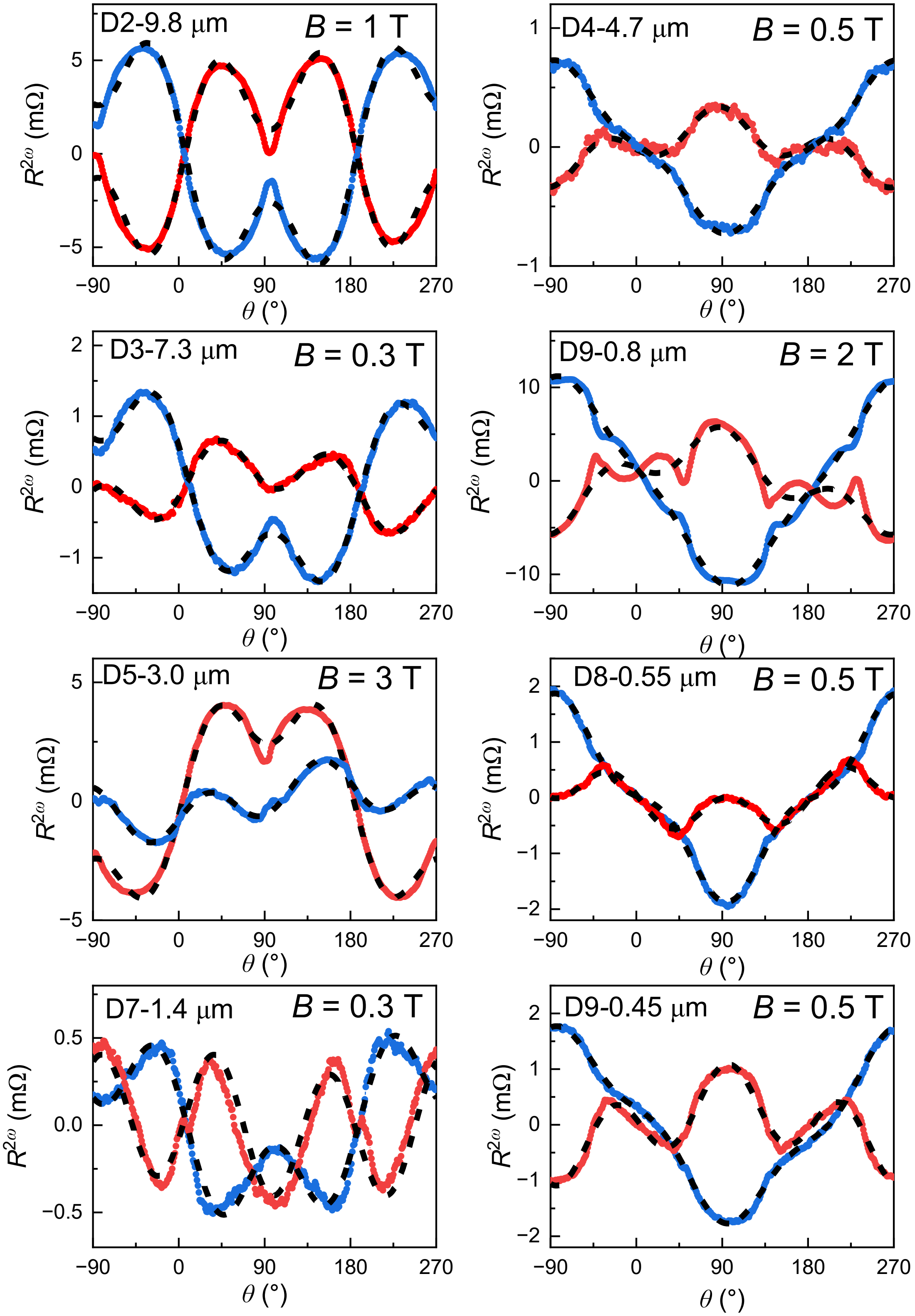

**Extended Data Fig. 10 Reproducibility of sin(3$\theta$) component in angular-dependent second-harmonic resistance $R^{2\omega}(\theta)$.** Higher order terms up to the sin(3$\theta$) term (black dashed lines) provide a good fit to the $R^{2\omega}(\theta)$ data from both the top (blue) and bottom (red) surfaces across all devices. The left and right panels show type-A and type-B patterns, respectively, showing a 50:50 statistical distribution.